\documentclass[%
 reprint,
superscriptaddress,
 amsmath,amssymb,
 aps,
prb,
]{revtex4-1}

\usepackage{graphicx}
\usepackage{dcolumn}
\usepackage{bm}
\usepackage{xcolor}

\begin{document}

\preprint{APS/123-QED}

\title{Interacting spin and charge density waves in kagome metal FeGe}

\author{Mason L. Klemm}
\email{masonklemm@gmail.com}
\author{Tingjun Zhang}
\affiliation{Department of Physics and Astronomy, Rice University, Houston, Texas, USA.}
\affiliation{Rice Laboratory for Emergent Magnetic Materials and Smalley-Curl Institute, Rice University, Houston, Texas, USA.}
\author{Barry L. Winn}
\author{Fankang Li}
\author{Feng Ye}
\affiliation{Neutron Scattering Division, Oak Ridge National Laboratory, Oak Ridge, Tennessee 37831, USA}
\author{Sijie Xu}
\author{Xiaokun Teng}
\author{Bin Gao}
\author{Ming Yi}
\affiliation{Department of Physics and Astronomy, Rice University, Houston, Texas, USA.}
\affiliation{Rice Laboratory for Emergent Magnetic Materials and Smalley-Curl Institute, Rice University, Houston, Texas, USA.}
\author{Pengcheng Dai}
 \email{pdai@rice.edu}
\affiliation{Department of Physics and Astronomy, Rice University, Houston, Texas, USA.}
\affiliation{Rice Laboratory for Emergent Magnetic Materials and Smalley-Curl Institute, Rice University, Houston, Texas, USA.}

\date{\today}

\begin{abstract}
Unveiling the interplay between spin density wave (SDW) and charge density wave (CDW) orders in correlated electron materials is important to obtain a comprehensive understanding of their electronic, structural, and magnetic properties. Kagome lattice materials are interesting because their flat electronic bands, Dirac points, and van Hove singularities can enable a variety of exotic electronic and magnetic phenomena. The kagome metal FeGe (the B35 phase), which exhibits a CDW order deep within an A-type antiferromagnetic (AFM) phase, was found to respond dramatically to post-growth annealing --- with the ability to tune the CDW repeatedly from long-range order to no (or extremely weak)  order. Additionally, neutron scattering studies suggest that incommensurate magnetic peaks that onsets at $T_{\rm Canting}=T_{\rm SDW} \approx 60$ K in the system arise from a SDW order instead of the AFM double cone structure. Here we use inelastic neutron scattering to show two distinct spin excitations exist below $T_{\rm Canting}$ corresponding to two coexisting magnetic orders in the system in both sets of annealed samples with and without CDW. While CDW order or no (or extremely weak)
order can dramatically affect the onset temperature of  $T_{\rm Canting}$ and elastic incommensurate magnetic scattering, its impact on low-energy spin fluctuations is more limited. In both samples, a pair of gapless incommensurate spin excitations arising from the SDW order wavevector coexist with gapped commensurate spin waves from the A-type AFM order across $T_{\rm Canting}$. The low-energy spin excitations for both samples couple dynamically to the lattice through enhanced magnetic scattering intensity on cooling below $T_{\rm CDW}$, regardless the status of the static long-range CDW order.
The incommensurate SDW order in the long-range CDW ordered sample also induces a tiny in-plane lattice distortion of the kagome lattice that is absent in the no (or extremely weak) CDW ordered sample, in a way different from the previously known SDW and CDW ordering materials. 
\end{abstract}

\maketitle


\section{\label{sec:level1} Introduction}
In most three-dimensional magnetic insulators, the magnetic order and associated excitations can be well described by quantum spin model with local moments on each atomic site without the involvement of the underlying lattice \cite{WOS:000207403700002,10.1093/oso/9780198862314.001.0001}. The Hamiltonian that describes these systems is the Heisenberg model, where the magnetic ordering temperature is determined by the competition between the thermal energy and the magnetic exchange interactions, and spin waves are the transverse motion of spins perpendicular to the ordered moment direction  \cite{WOS:000207403700002,10.1093/oso/9780198862314.001.0001}. 
 In the magnetic ordered state, spin waves must obey the detailed balance through the Bose population factor \cite{WOS:000207403700002,10.1093/oso/9780198862314.001.0001}. On warming to the magnetic disordered state, 
 spins fluctuate dynamically and incoherently, and 
 paramagnetic spin fluctuations (excitations) should be featureless in momentum space and decrease with increasing temperature  \cite{WOS:000207403700002,10.1093/oso/9780198862314.001.0001}.
For metallic magnets, magnetic order may arise from quasiparticle spin-flip excitations between the valence (hole) and conduction (electron) bands at the Fermi level as a consequence of electron-electron correlations \cite{RevModPhys.66.1}. This so-called spin density wave (SDW) order is usually accompanied or preceded by a charge density wave (CDW) order associated with a lattice distortion as seen in pure Cr metal \cite{RevModPhys.60.209,RevModPhys.66.25,PhysRevB.51.10336}, hole-doped copper oxides \cite{RevModPhys.75.1201,annurev:/content/journals/10.1146/annurev-conmatphys-032922-094430}, and iron pnictides \cite{RevModPhys.87.855}.In the case of Cr, SDW and CDW orders occur simultaneously below $T_{\rm SDW}$ at related incommensurate wavevectors \cite{PhysRevB.51.10336}, thus establishing the SDW-CDW coupling \cite{RevModPhys.60.209,RevModPhys.66.25}.

Kagome metals are an exciting platform to study a host of exotic electronic and magnetic phenomena including superconductivity \cite{ortiz_cs_2020,WOS:001221002000001}, time reversal symmetry breaking \cite{ye_massive_2018}, CDW order \cite{PhysRevLett.129.216402,doi:10.1021/jacs.4c16347}, and quantum spin liquids \cite{broholm_quantum_2020}. Such phenomena arise from the unique lattice geometry of corner-sharing triangles in kagome metals that yield flat electronic bands in a close proximity to the Fermi surface, discontinuities like van Hove singularities and Dirac cones, and magnetic frustration in certain magnetically ordered configurations \cite{kang_topological_2020,WOS:000934065100013}. Recently, CDW order in kagome metals has been extensively investigated in the nonmagnetic superconducting $A$V$_3$Sb$_5$ ($A=$ K, Rb, Cs) \cite{ortiz_cs_2020,WOS:001221002000001}, nonmagnetic nonsuperconducting ScV$_6$Sn$_6$ \cite{PhysRevLett.129.216402} and $Ln$Nb$_6$Sn$_6$ ($Ln=$ Ce-Lu, Y) \cite{doi:10.1021/jacs.4c16347}, and magnetic nonsuperconducting FeGe \cite{doi:10.1143/JPSJ.18.589,beckman_susceptibility_1972,bernhard_magnetic_1988,teng_discovery_2022,PhysRevLett.129.166401}. Magnetic FeGe is particularly interesting because CDW order $T_{\rm CDW}=100$ K was discovered deep within the A-type antiferromagnetic (AFM) order phase $T_{\rm N}=410 $ K --- a novel phenomena in CDW and magnetic ordered materials (Fig. 1a) \cite{beckman_susceptibility_1972,bernhard_magnetic_1988,teng_discovery_2022}. Furthermore, the short-range CDW in as-grown FeGe can be tuned by post-growth annealing to achieve both long-range CDW and extinguished (or extremely weak) CDW order upon successive annealing cycles \cite{wu_annealing-tunable_2024,WOS:001276353800013,klemm_vacancy-induced_2025}. 

The crystal structure of kagome lattice FeGe is the CoSn type in the $P6/mmm$ space group with two Ge positions (Fig. 1a, Ge1 in the Fe$_3$Ge layer and Ge2 between the Fe$_3$Ge layers) \cite{doi:10.1143/JPSJ.18.589,beckman_susceptibility_1972,bernhard_magnetic_1988}. It has an A-type AFM structure where $c$-axis polarized Fe moments within each kagome layer are aligned antiferromagnetically along the $c$-axis below a N$\rm \acute{e}$el temperature $T_{\rm N}\approx 410$ K \cite{beckman_susceptibility_1972,bernhard_magnetic_1988}. Below $\sim$100 K a $2\times 2 \times2$ CDW order emerges, structurally similar to that in CsV$_3$Sb$_5$ \cite{teng_discovery_2022,WOS:001164031000002,WOS:001221002000001}. However, unlike the CDW in $A$V$_3$Sb$_5$, which is believed to arise from Fermi surface nesting \cite{li_observation_2021}, CDW in FeGe is theorized to originate from dimerization of Ge1 atoms along the $c$-axis as a magnetic energy saving measure \cite{wang_enhanced_2023,PhysRevB.108.035138,PhysRevResearch.6.033222}. A swath of spectroscopic experimental evidence has emerged in support of the dimerization picture in FeGe \cite{wu_annealing-tunable_2024,WOS:001276353800013,chen_discovery_2024,klemm_vacancy-induced_2025,Subires_2025,Shi_2024}, while angle-resolved photoemission spectroscopy measurements cast severe doubt on a Fermi surface nesting origin \cite{teng_magnetism_2023,zhao_photoemission_2023,oh_tunability_2024}. FeGe exhibits an additional magnetic transition below $T_{\rm Canting}\approx 60$ K, conventionally understood as a double-cone AFM structure with small canting from the $c$-axis, producing incommensurate magnetic peaks along the $c$-axis at $q_{IC}=(L\pm\delta)$, where $\delta\approx0.04$ reciprocal lattice units (r.l.u.) and $L=\pm1/2,3/2,\cdots$ (Fig. 1b pink dots) \cite{bernhard_magnetic_1988}. The incommensurate peak intensities are modified dramatically by the presence and absence of long-range CDW, indicating a strong coupling between CDW and incommensurate magnetic order in the system \cite{klemm_vacancy-induced_2025}. While an anomalous Hall effect (AHE) was found to onset approximately below $T_{\rm CDW}$ in as-grown FeGe \cite{teng_discovery_2022}, the AHE in annealed long-range CDW samples is enhanced by an order of magnitude, on par with the giant AHE observed in KV$_3$Sb$_5$ \cite{klemm_vacancy-induced_2025,yang_giant_2020}, and appears below $T_{\rm Canting}$ \cite{klemm_vacancy-induced_2025}. Since our elastic \cite{klemm_vacancy-induced_2025} and inelastic \cite{chen_competing_2024} neutron scattering experiments suggest that the incommensurate magnetic peaks arise from a SDW instead of a AFM double-cone structure \cite{bernhard_magnetic_1988}, it will be important to determine how incommensurate order and associated spin excitations are affected by the formation of 
a long range
CDW order, and compare the outcome with other SDW-CDW coexisting materials \cite{RevModPhys.60.209,RevModPhys.66.25,PhysRevB.51.10336,RevModPhys.75.1201,annurev:/content/journals/10.1146/annurev-conmatphys-032922-094430,RevModPhys.87.855}.

The A-type AFM order is conventionally thought to cant the local $c$-axis Fe moment towards the kagome plane to form a double-cone magnetic structure at $T_{\rm Canting}\approx 60$ K \cite{teng_discovery_2022,bernhard_magnetic_1988}. However, recent magnetic refinement of post-growth annealed samples via neutron scattering is inconsistent with the magnetic structure factor predicted from a double-cone model \cite{klemm_vacancy-induced_2025}. Additionally, inelastic neutron scattering experiments on as-grown FeGe single crystals reveal gapless incommensurate spin excitations, persisting at temperatures well above the static incommensurate order,  that merge into a gapped commensurate excitation at 1 meV \cite{chen_competing_2024}. The persistence of the incommensurate excitations above the canting temperature is incompatible with a simple canting of the A-type magnetic order,  suggesting it being a SDW order.

In this work, we report elastic and inelastic neutron scattering experiments on two sets of post-growth annealed samples with long-range CDW order and no (or extremely weak) CDW order \cite{klemm_vacancy-induced_2025}. From neutron diffraction measurements of long-range CDW and no
(or extremely weak)
CDW ordered samples, we find that incommensurate magnetic Bragg peaks in the no
(or extremely weak)
CDW sample are suppressed dramatically compared with the long-range CDW sample, but the suppression differs along the $c$-axis and in-plane directions possibly due to directional dependent magnetic form factors of the SDW order.
Although our inelastic neutron scattering measurements on both samples reveal gapless incommensurate spin excitations stemming from the elastic positions \cite{chen_competing_2024}, the incommensurate spin excitations extend beyond 1 meV and coexist with gapped commensurate spin waves stemming from the A-type AFM order, consistent with the SDW excitations coexisting with the A-type local moment spin waves. We also carried out neutron Larmor diffraction measurements to study the spin-lattice coupling in long-range and no (or extremely weak) CDW ordered FeGe \cite{li_high_2017,PhysRevB.93.134519,wu_symmetry_2024}. The neutron Larmor diffraction measurements further reveal that the SDW order is coupled with the lattice distortion of hexagonal kagome plane in the long-range CDW sample but is absent in
the no (or extremely weak) CDW sample, different from the usual SDW-CDW coupling in Cr  \cite{RevModPhys.60.209,RevModPhys.66.25,PhysRevB.51.10336}, hole-doped copper oxides \cite{RevModPhys.75.1201,annurev:/content/journals/10.1146/annurev-conmatphys-032922-094430}, and iron pnictides \cite{RevModPhys.87.855}.  
  
\begin{figure}
    \centering
    \includegraphics[width=0.9\linewidth]{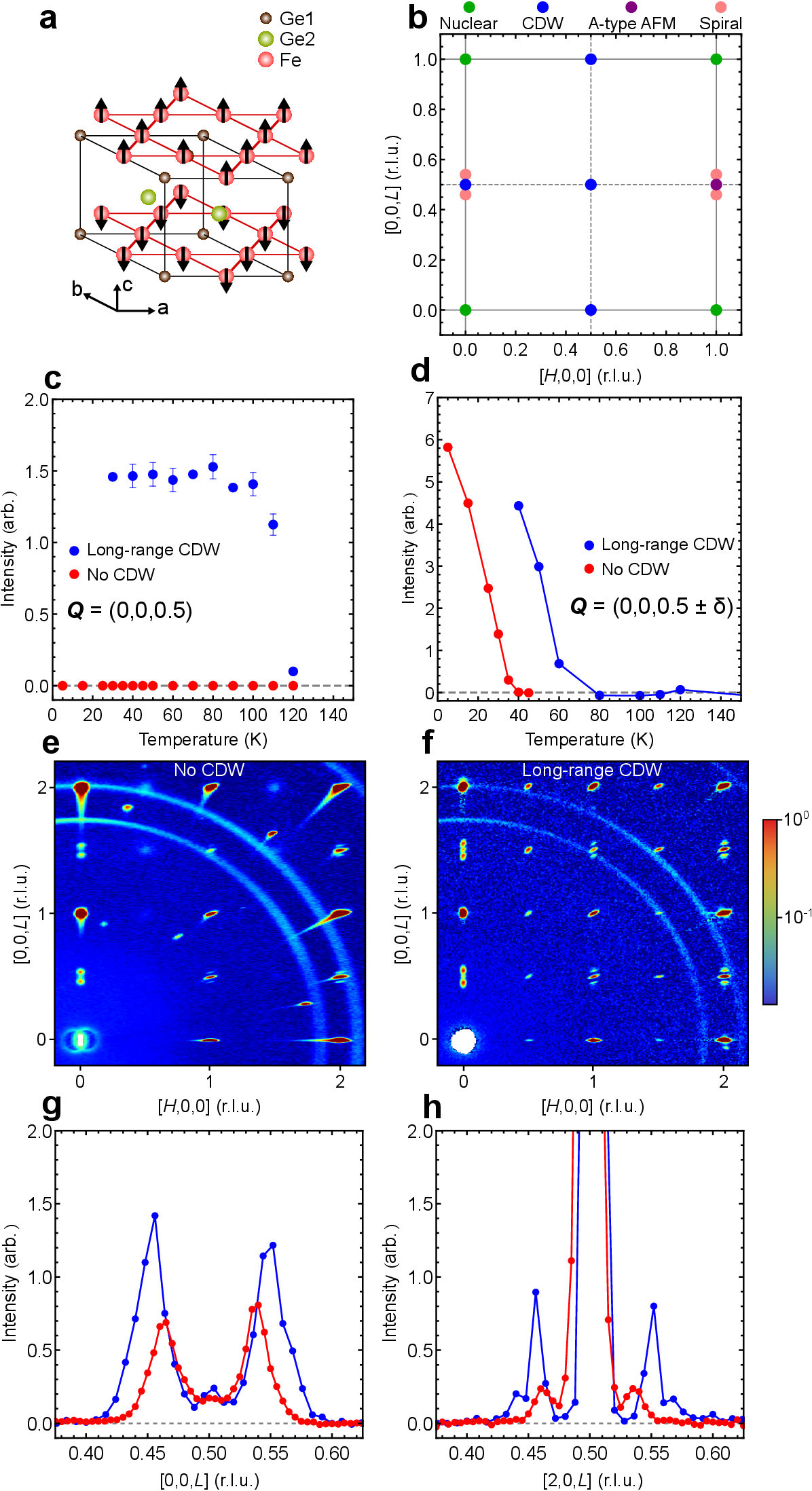}
    \caption{\textbf{a.} A-type antiferromagnetic order on the FeGe kagome lattice. Two Ge positions are identified as Ge1 and Ge2. \textbf{b.} Schematic of reciprocal space showing the locations of various CDW, nuclear, commensurate and incommensurate magnetic Bragg peaks at base temperature in long-range CDW ordered samples. \textbf{c.} Order parameter scan of the CDW Bragg peak $\textbf{Q} = (0,0,0.5)$ for samples with no CDW order and long-range CDW order. \textbf{d.} Order parameter scan of the incommensurate magnetic Bragg peaks $\textbf{Q} = (0,0,0.5\pm \delta)$ for both sample types.  \textbf{e-f.} Reciprocal space map in log scale of sample with no CDW and long-range CDW order respectively at 5 K. Peaks not located at integer or half-integer in \textbf{e} are a result of a second grain within the single crystal and can be ingored, as these peaks randomly coincide with the $[H,0,L]$ plane of the primary aligned crystal grain. \textbf{g-h.} \textbf{Q}-cuts of the incommensurate magnetic Bragg peaks around $\textbf{Q}=(0,0,0.5)$ and $(2,0,0.5)$ respectively comparing long-range CDW samples (blue) and no CDW samples (red). Reciprocal space map intensities, and by consequence \textbf{Q}-cuts, of both annealed samples are normalized to the $\textbf{Q} = (0,0,1)$ nuclear Bragg peak. Error bars correspond to one standard deviation.}
    \label{fig:enter-label}
\end{figure}

\section{Results and Discussion}
For our elastic and inelastic neutron scattering experiments, roughly three grams of FeGe single crystals were grown via chemical vapor transport as described in \cite{teng_discovery_2022}. Half of the samples were annealed at 320$^\circ$C for 96 hours yielding long-range CDW ordered samples. The other half were annealed at 560$^\circ$C for 96 hours to obtain samples with no (or extremely weak) CDW order \cite{klemm_vacancy-induced_2025}. The two sets of annealed samples were co-aligned on two different thin sheets of aluminum in the $[H, 0,L]$ scattering plane for inelastic neutron scattering experiments carried out on HYSPEC neutron time-of-flight spectrometers at the Spallation Neutron Source (SNS), Oak Ridge National Laboratory (ORNL) \cite{refId0}. For diffraction measurements, we used one piece of single crystal of FeGe mounted on an aluminum pin using the CORELLI, BL-9, spectrometer \cite{Ye:ut5001} of the SNS, ORNL. The momentum transfer $\textit{\textbf{Q}}$ in three dimensional reciprocal space in \AA$^{-1}$ was defined as $\textit{\textbf{Q}}= H {\textbf{a}}^\ast +  K {\textbf{b}}^\ast +  L {\textbf{c}}^\ast$, where $H$, $K$, and $L$ are Miller indices and ${\textbf{a}}^\ast=2\pi ({\textbf{b}}\times {\textbf{c}})/[{\textbf{a}}\cdot ({\textbf{b}}\times {\textbf{c}})]$, ${\textbf{b}}^\ast=2\pi ({\textbf{c}}\times {\textbf{a}})/[{\textbf{a}}\cdot ({\textbf{b}}\times {\textbf{c}})]$, ${\textbf{c}}^\ast=2\pi ({\textbf{a}}\times {\textbf{b}})/[{\textbf{a}}\cdot ({\textbf{b}}\times {\textbf{c}})]$, with $\textbf{a}=a\ {\bf \hat{x}}$, $\textbf{b}=a (\cos{120}\ {\bf \hat{x}}+\sin{120}\ {\bf \hat{y}} )$, and $\textbf{c}=c\ {\bf \hat{z}}$ ($a\approx b\approx 4.99$ \AA, $c\approx 4.05$ \AA\ at room temperature).

We first compare neutron diffraction results on FeGe with long-range CDW \cite{klemm_vacancy-induced_2025} and no (or extremely weak) CDW order. Order parameter scans performed on the CDW Bragg peak $\textit{\textbf{Q}} = (0,0,0.5) $ reveal an onset of $T_{\rm CDW}=120$ K for CDW order in samples with long-range order, and no signal at the CDW wavevector $\textit{\textbf{Q}} = (0,0,0.5)$ in samples with no CDW (Fig. 1c). Order parameter scans of the incommensurate magnetic order at  $\textit{\textbf{Q}} = (0,0,0.5\pm \delta) $ where $\delta=0.04$ show a suppression of SDW order for samples with no  CDW and an enhancement of SDW order in samples with long-range CDW order, consistent with recent neutron measurements (Fig. 1d) \cite{klemm_vacancy-induced_2025}. Figures 1e,f show an overall map of the reciprocal space within the $[H,0,L]$ scattering plane for no CDW and long-range CDW samples, respectively, at 5 K in a log scale.

Although we cannot detect any static long-range CDW order within the sensitivity of our measurements, extremely weak and diffuse scattering exists at the CDW wavevectors in samples termed no CDW (Fig. 1e). We emphasize the phrasing ``no CDW" is useful in delineating the broad spectrum of CDW tunability in the system, as important electronic signatures --- including a kink in resistivity and magnetic susceptibility at $T_{\text{CDW}}$ and an absence of AHE --- are absent. It is likewise important to note the reciprocal space map in Fig. 1e is the product of one single crystal whereas the order parameter in Fig. 1c is taken from Fig. 3g of co-aligned samples during our inelastic neutron scattering measurements. 

Table I shows the magnetic intensity ratio of commensurate magnetic Bragg peaks for no (or extremely weak) CDW and long-range CDW ordered samples, suggesting both samples having a simple A-type magnetic structure. To determine the impact of CDW order on incommensurate magnetic peaks, we compare in Table II the ratio of incommensurate peaks for no CDW and long-range CDW ordered samples. If the presence of CDW order uniformly modifies the incommensurate magnetic order, we would expect an equivalent intensity ratio for all $\textit{\textbf{Q}}$ as seen for commensurate magnetic peaks shown in Table I. However, we find that magnetic scattering intensity for FeGe without CDW order is suppressed much more dramatically along the $[H,0,\pm 0.5\pm\delta]$ direction compared with peaks along the $[0,0,L\pm\delta]$ direction (Table II). While these results indicate that CDW order has a clear impact of magnetic scattering intensity of the SDW order, it is unclear why the scattering intensity reduction behaves differently along the $[H,0,\pm 0.5\pm\delta]$ and $[0,0,L\pm\delta]$ directions. 

One possibility is that the magnetic form factor of incommensurate peaks is impacted by the presence of CDW in the system. In isolation, the magnetic form factor of a magnetic ion is determined by the spatial distribution of its electron spin and orbital magnetic moment, and can be used to characterize magnetic signals in the vast majority of solids. However, if incommensurate magnetic order in FeGe arises from Fermi surface nesting of hole and electron Fermi pockets, its magnetic form factor may be anisotropic and couple with details of Fermi surface nesting and the onset of CDW order. A detailed investigation is beyond the scope of this work.

\begin{table}[]
\centering
\begin{tabular}{ccccc}
\hline
$H$ & $K$ & $L$   & Ratio (No/long-range CDW) & Errors  \\ \hline
1 & 0 & 0.5 & 1.28            & 0.03     \\
2 & 0 & 0.5 & 1.19            & 0.02     \\
2 & 0 & 1.5 & 1.11            & 0.03     \\
2 & 0 & 2.5 & 1.08            & 0.07     \\
3 & 0 & 0.5 & 1.18            & 0.19     \\
4 & 0 & 0.5 & 1.23            & 0.10     \\
4 & 0 & 1.5 & 1.12            & 0.15     \\
4 & 0 & 2.5 & 1.13            & 0.28     \\ \hline
\end{tabular}
\caption{Integrated intensity ratio of A-type commensurate magnetic Bragg peaks for No CDW and long-range CDW ordered samples. The integrated magnetic scattering intensity for long-range CDW ordered sample are obtained by using data at 135 K without the complication of CDW order. Those for no CDW sample are integrated magnetic scattering at 55 K.}
\end{table}

\begin{table}[]
\centering
\begin{tabular}{ccccc}
\hline
$H$ & $K$ & $L$     & Ratio (No/long-range CDW) & Errors \\ \hline
0 & 0 & 0.455 & 0.245           & 0.004    \\
0 & 0 & 0.545 & 0.297           & 0.005    \\
0 & 0 & 1.455 & 0.27            & 0.01     \\
0 & 0 & 1.545 & 0.22            & 0.01     \\
0 & 0 & 2.455 & 0.24            & 0.03     \\
0 & 0 & 2.545 & 0.20            & 0.03     \\ \hline
1 & 0 & 0.455 & 0.13            & 0.02     \\
2 & 0 & 0.455 & 0.16            & 0.01     \\
2 & 0 & 0.545 & 0.16            & 0.01     \\
2 & 0 & 1.545 & 0.12            & 0.02     \\ \hline
\end{tabular}
\caption{Integrated intensity ratio of incommensurate magnetic Bragg peaks for No CDW and long-range CDW ordered samples.
The intensity reduction at ${\bf Q}=(0,0,L\pm\delta)$ ($L=0.5, 1.5, 2.5$) due 
to No CDW is shown in the top half of the table, while those at ${\bf Q}=(H,0,L\pm\delta)$ ($H=1,2$ and $L=0.5, 1.5$) are at the bottom
half of the table. 
}
\end{table}

Figures 2a-d summarize the energy and momentum dependent low-energy spin excitations $S(\textit{\textbf{Q}},E)$ in the long-range and no (or extremely weak) CDW ordered FeGe at temperatures above $T_{\rm Canting}$. The low-energy spin excitations along the $[0,0,L]$ direction reveal two distinct excitations in both samples with long-range and no (or extremely weak)
CDW order (Figs. 2a-d). Below 1 meV two gapless incommensurate excitations exist at  $\textit{\textbf{Q}} = (0,0,0.5\pm \delta)$ where $\delta = 0.04$, stemming from elastic positions and consistent with previous inelastic neutron scattering results \cite{chen_competing_2024}. However, above 1 meV a gapped commensurate peak coexists with the gapless incommensurate peaks, as revealed by \textit{\textbf{Q}}-cuts at 1.2 meV for both long-range CDW and no (or extremely weak) CDW samples (Figs. 2b and 2d). Previously, it was believed that the incommensurate excitations merge into a single commensurate signal above 1 meV but our results show conclusively that below the incommensurate magnetic order transition there exist two coexisting orders: an A-type antiferromagnetic order responsible for the gapped commensurate signal and SDW order accounting for the gapless incommensurate signal \cite{chen_competing_2024}. 

\begin{figure}
     \centering
   \includegraphics[width=0.9\linewidth]{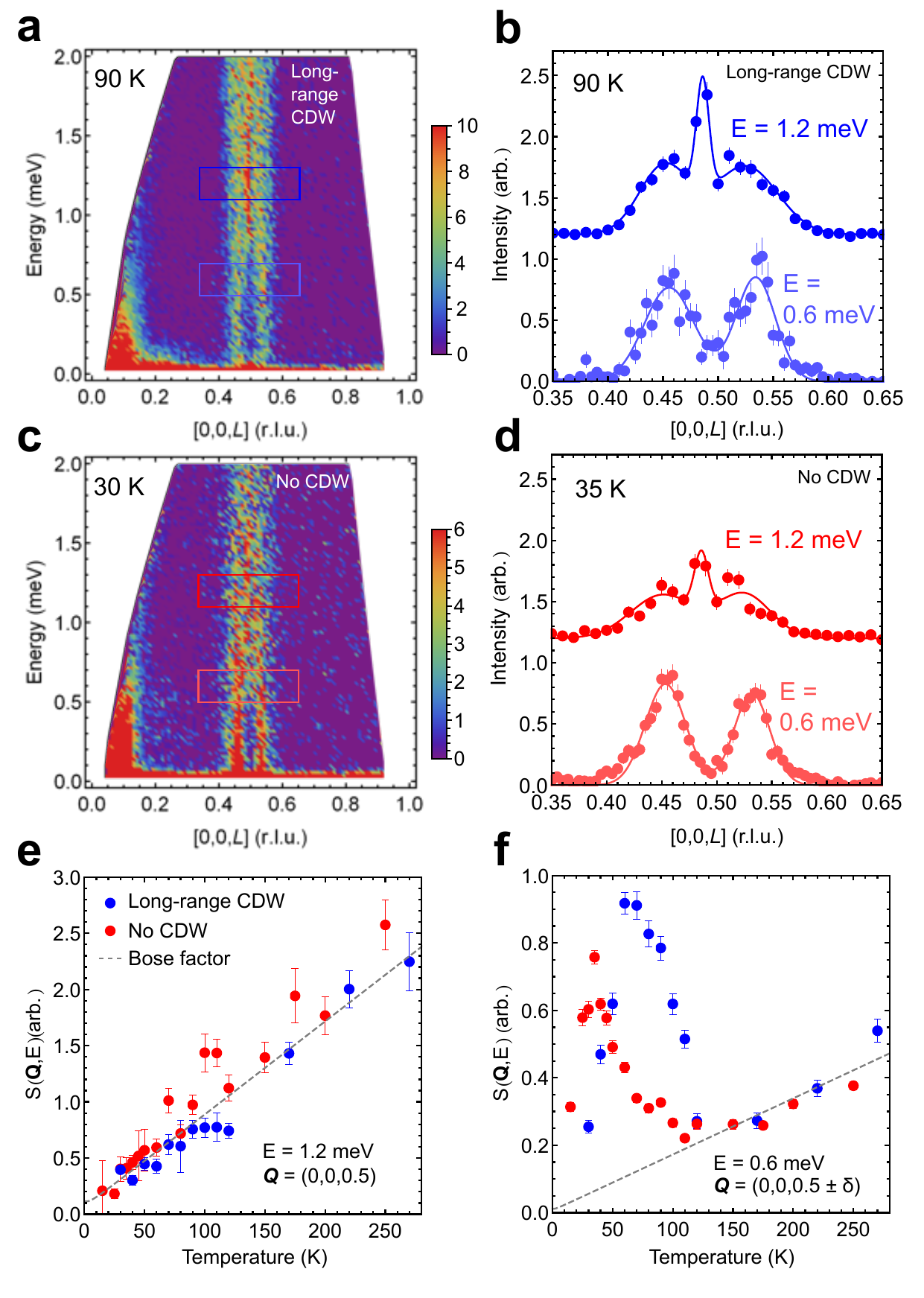}
     \caption{\textbf{a.} Low energy excitation spectra of the long-range CDW ordered sample at 90 K. \textbf{b.} Constant energy cuts along $\textbf{Q} = [0,0,L]$ using the integrated intensity range depicted by rectangles in \textbf{a}. \textbf{c-d.} Same as \textbf{a-b} but for samples with no CDW order. \textbf{e.} Commensurate excitation intensity as a function of temperature extracted from the three-gaussian fits to the \textbf{Q}-cuts at 1.2 meV in \textbf{b,d} for samples with and without long-range CDW order. \textbf{f.} Incommensurate excitation intensity as a function of temperature extracted from the three-gaussian fits to the \textbf{Q}-cuts at 0.6 meV in \textbf{b,d}. The dashed line in \textbf{e-f} is the Bose population factor. Error bars correspond to one standard deviation.}
    \label{fig:enter-label}
  \end{figure}

To separate commensurate spin waves from the A-type AFM order from incommensurate spin excitations at 1.2 meV for the long-range and no (or extremely weak) CDW ordered FeGe, we fit the observed spin excitations spectra with three Gaussian peaks on a linear background and extrapolated commensurate spin wave intensity at $\textit{\textbf{Q}} = (0,0,0.5) $ (Figs. 2b and 2d). The temperature dependence of the commensurate spin excitation at 1.2 meV follows the Bose population factor, given by 
\begin{equation}
    I  \propto \frac{1}{1-e^{\frac{-E}{k_BT}}}
\end{equation}

in both samples (Fig. 2e). For direct comparisons of signals between the long-range CDW and no (or extremely weak) CDW sample arrangements, the elastic nuclear Bragg peak $\textit{\textbf{Q}} = (1,0,0)$ intensities at 300 K are used to normalize the two samples to account for differences in mass and sample environment background. Notably, the intensity and evolution of the commensurate excitation with temperature are effectively identical, reflecting the substantial invariance of the A-type antiferromagnetic order under different annealing conditions found recently \cite{klemm_vacancy-induced_2025}. A comparison of the temperature dependence of the incommensurate spin excitation at 0.6 meV shows agreement with the Bose population factor above the CDW transition, but deviates from the Bose population factor below the CDW transition (Fig. 2f). The incommensurate peak intensity was extracted from the two Gaussian peak fit shown in Figs. 2b and 2d. The long-range CDW and no (or extremely weak) CDW incommensurate inelastic signals peak at their respective SDW transitions of $\sim$80 K and 35 K, consistent with the second-order magnetic phase transitions (Fig. 2f). The dashed line is the expected temperature-dependent $S(\textit{\textbf{Q}},E)$ with the Bose population factor, clearly inconsistent with the data \cite{chen_competing_2024}.

\begin{figure}
     \centering
     \includegraphics[width=1\linewidth]{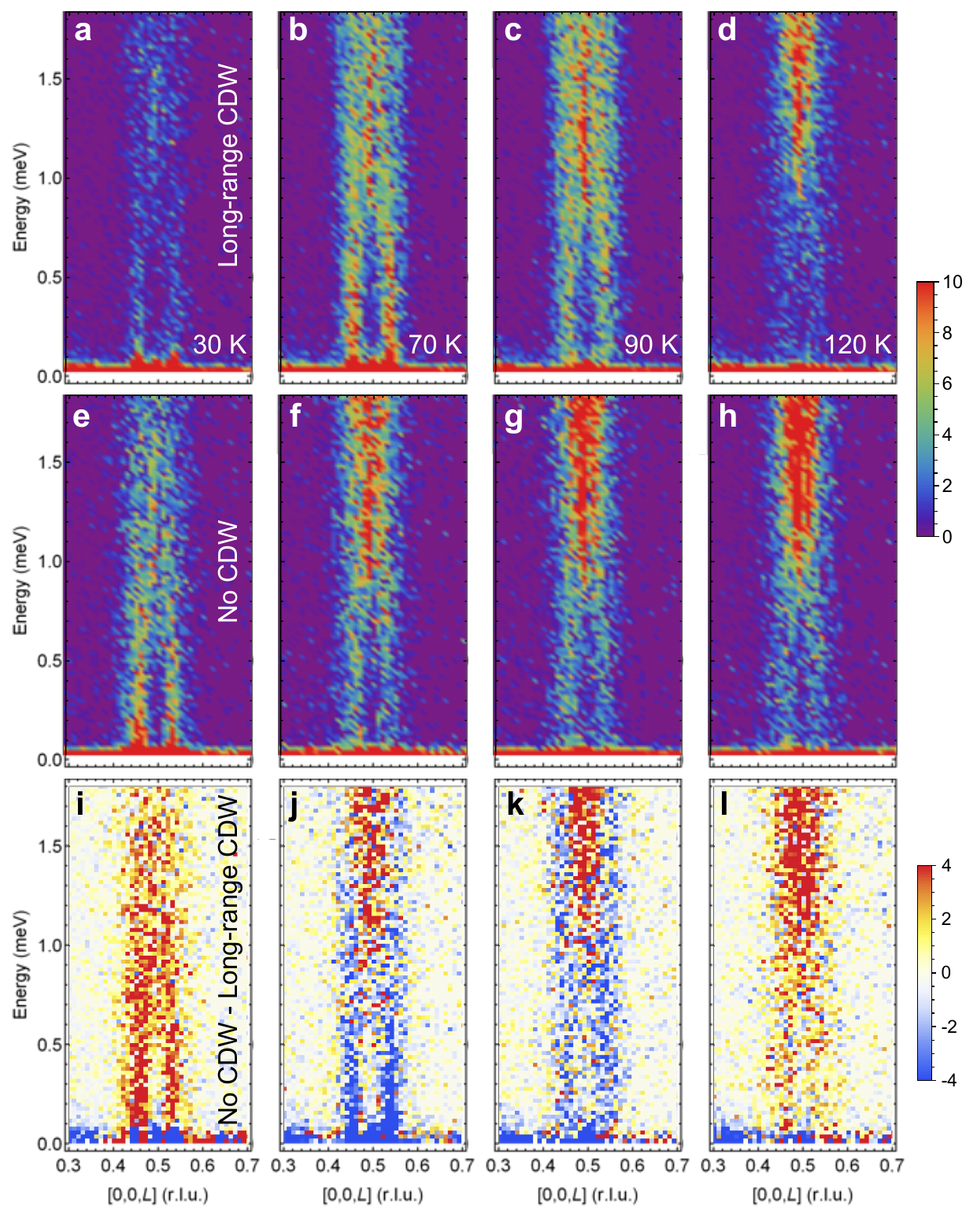}
 \caption{\textbf{a-d.} Low energy excitation spectra along $\textbf{Q}=[0,0,L]$ for long-range CDW ordered samples at 30, 70, 90, and 120 K respectively. \textbf{e-h.} Same as \textbf{a-d}, but for samples with no CDW order. \textbf{i-l.} Difference of \textbf{e-h} and \textbf{a-d} respectively. The excitation spectra of no CDW samples and long-range CDW ordered samples are normalized to the nuclear Bragg peak $\textbf{Q}=(1,0,0)$ and background subtracted to appropriately take the difference in \textbf{i-l}.}
     \label{fig:enter-label}
 \end{figure}

The raw low-energy spin excitation data at key temperatures 30 K, 70 K, 90 K, and 120 K are shown in Figures 3a-3h for both samples. The 30 K data corresponds to the maximum in the incommensurate signal of the no (or extremely weak) CDW sample (Fig. 3e), 70 K corresponds to the maximum in the incommensurate signal of the long-range CDW ordered sample (Fig. 3b). The 90 K (Figs. 3c and 3g) and 120 K (Figs. 3d and 3h) data are below and above the CDW transition, respectively. Since commensurate spin waves from A-type order are expected to behave similarly for both the long-range and no (or extremely weak) CDW ordered samples, the difference between the no (or extremely weak) CDW and long-range CDW samples using normalized and background subtracted data highlights the different evolution of spin excitations with temperature (Figs. 3i-3l). The red regions represent a dominant contribution from no (or extremely weak)
CDW samples and likewise blue regions for long-range CDW samples (Figs. 3i-3l). The elastic incommensurate signal at $\textit{\textbf{Q}} = (0,0,0.5\pm \delta)$ of the long-range CDW sample is dominant at 30 K and 70 K (Figs. 3i,3j) before disappearing above the SDW transition, around 70 K (Fig. 3k,l), as illustrated in the order parameter scan in Fig. 1c. Similarly, the elastic CDW peak at $\textit{\textbf{Q}} = (0,0,0.5)$ can be seen up to 90 K (Figs. 3i,3k) from the long-range CDW sample, reflecting the order parameter in Fig. 1b. The gapless incommensurate excitations of both samples experience a trade off of intensity as they approach their respective SDW transition temperatures. At 30 K the incommensurate excitation of the no 
(or extremely weak)
CDW sample dominates (Fig. 3i) while at 70 K and 90 K the long-range CDW sample dominates (Fig. 3j,3k) before both samples have roughly equivalent excitation intensity at 120 K (Fig. 3l). 

A clear illustration of the two distinct excitations is in Fig. 3k, where the dominant incommensurate excitation of the long-range CDW sample persist to 1.8 meV while the slightly larger commensurate signal from the no 
(or extremely weak)
CDW sample is clearly distinct from 1 meV to 1.8 meV. The linear increase of the gapped commensurate excitations above 1 meV for both samples can be seen in the raw data (Figs. 3a-3h). The subtracted data (Figs. 3i-3l) appears to suggest the no
(or extremely weak)
CDW commensurate excitation dominates at all temperatures, but both fitted intensities at 1.2 meV follow the Bose population factor within error bars (Fig. 2a). We emphasize that the selected temperatures in Fig. 3 correspond to temperatures where the commensurate excitation intensity happens to be higher for no (or extremely weak) CDW samples, however the larger trend of the commensurate excitation is effectively equivalent for both samples (Fig. 2a).     

\begin{figure}
    \centering
    \includegraphics[width=1\linewidth]{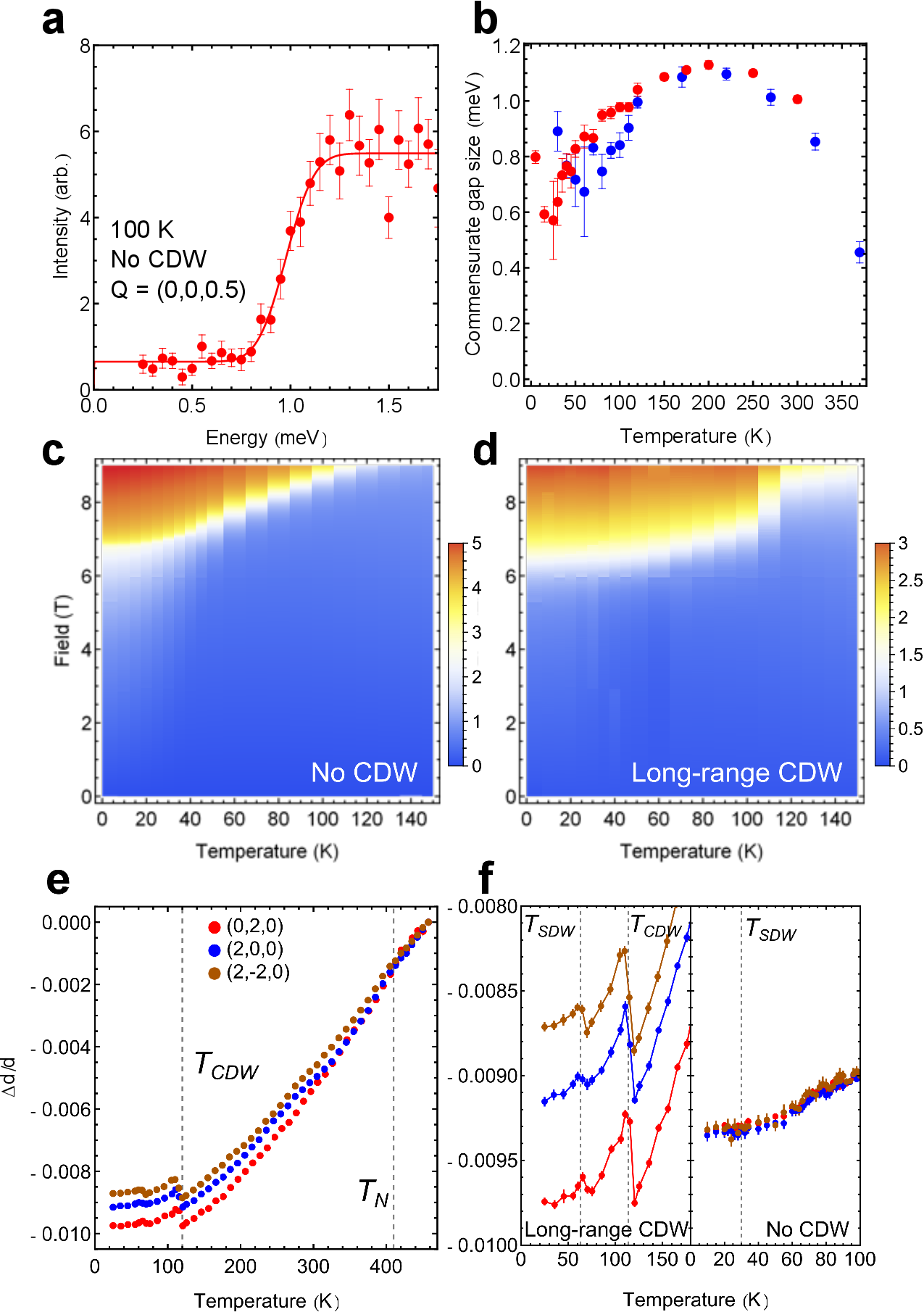}
    \caption{\textbf{a.} Sample fitting of the commensurate energy gap using Eq. 2 and integration range $\textbf{Q}=(0,0,0.5\pm 0.01)$. \textbf{b.} Commensurate energy gap size extracted from fits described in \textbf{a} as a function of temperature for no CDW samples (red) and long-range CDW samples (blue). \textbf{c-d.} Isothermal magnetic susceptibility heat map demonstrating the evolution of the spin flop transition in no CDW and long-range CDW samples respectively. \textbf{e.} Lattice spacing evolution of  $\textbf{Q}=(0,2,0),\ (2,0,0),\ \text{and}\ (2,-2,0)$ nuclear Bragg pea ks as a function of temperature using neutron Larmor diffraction in a long-range CDW sample. \textbf{f.} A zoomed in view of \textbf{e} showing spontaneous lattice expansion at $T_{SDW}$ and $T_{CDW}$, plus Larmor diffraction across $T_{SDW}$ in the sample with no CDW. Error bars represent one standard deviation.}
    \label{fig:enter-label}
\end{figure}

In a classic local moment system, one would only 
expect the intensity of spin waves to follow the Bose population factor in the magnetic ordered state \cite{WOS:000207403700002,10.1093/oso/9780198862314.001.0001}. While the temperature-dependent commensurate scattering at 1.2 meV follows the
expectation (Fig. 2e), the temperature-dependent incommensurate magnetic scattering at 0.6 meV for both samples show a considerable upturn below about 100 K and follows the Bose population factor for temperatures between 100 K and 250 K. These results suggest that regardless the status of the static long-range CDW order in FeGe, incommensurate spin fluctuations are coupled with the lattice dynamics associated with the static CDW order. If incommensurate magnetic order arises from a double cone local moment below $T_{\rm Canting}$ for both the
long-range and no (or extremely weak) CDW ordered samples \cite{beckman_susceptibility_1972,bernhard_magnetic_1988}, one would not expect the 0.6 meV 
paramagnetic spin excitations in these materials to couple with  
$T_{\text{CDW}}$ and 
obey the Bose population factor at temperatures above $T_{\text{CDW}}$ (Fig. 2f). However, if incommensurate magnetic order and associated 
spin fluctuations in FeGe arise from hole-electron Fermi surface nesting, where the nest 
condition is first established below $T_{\text{N}}$\cite{chen_competing_2024}, the incommensurate spin fluctuations can be regarded as imperfect nesting instead of uncorrelated paramagnetic scattering of a local moment system \cite{WOS:000207403700002,10.1093/oso/9780198862314.001.0001}. Since quasiparticle excitations from the hole-electron Fermi surface nesting are bosons, they must obey the detailed balance and the Bose population factor regardless if a long-range CDW order is established or not in FeGe.

Previously, the size of the commensurate gap was shown to evolve with the critical spin flop field \cite{chen_competing_2024}. Here we provide a more detailed look at the low temperature evolution of the spin-flop field and commensurate gap in post-growth annealed samples. A small integration range of $L = 0.5\pm0.01$ was used to extract the energy dependence from the raw data and was fit with the equation 
\begin{equation}
    I = I_0+\frac{A\cdot \text{Erfc}[(E_{gap}-E)/\sigma]}{1-e^{\frac{-E}{k_BT}}}
\end{equation}
to obtain the size of the gap, $E_{gap}$ (Fig. 4a). The gap shows a minimum at the SDW transition temperature in both samples and increases to 200 K before dropping, consistent with previous measurements (Fig. 4b) \cite{teng_discovery_2022,chen_competing_2024}. A heat map of the isothermal magnetization capturing the spin flop transition highlights the similarity of the spin flop critical field and commensurate gap size in both samples (Figs. 4c,4d). The sample with no (or extremely weak) CDW has a low temperature minimum in the spin-flop transition and a gradual evolution across the CDW transition (Fig. 4c) whereas the long-range CDW sample has a critical field minimum at higher temperatures and an abrupt increase at the CDW transition temperature (Fig. 4d), mirroring the behavior of the energy gap in Fig. 4b. Such behavior is expected, as the Zeeman energy $g\mu _{B}H$ is field dependent and induces a spin flop transition when it exceeds $E_{gap}$.

In previous work on as-grown FeGe, neutron Larmor diffraction was utilized to identify small structural distortions of the kagome lattice as a function of temperature on the order of $\Delta d/d \sim (10^{-6})$ \cite{wu_symmetry_2024}. Neutron Larmor diffraction is a polarized neutron scattering technique that exploits the Larmor precession of the neutron's dipole moment as it travels through a small static magnetic guide field  \cite{li_high_2017,PhysRevB.93.134519}. As the lattice shrinks or expands, the travel distance of the neutron changes and modifies the Larmor phase seen by the detector. The Larmor phase shift $\Phi$ is linearly related to the lattice spacing $d$ by $\Delta \Phi/\Phi = \Delta d/d$. More technical descriptions of the neutron Larmor diffraction method can be found in Refs. \cite{li_high_2017,PhysRevB.93.134519,wu_symmetry_2024}. We examined the lattice spacing of three equivalent Bragg peaks within the $P6/mmm$ space group, $\textbf{Q}=(0,2,0),\ (2,0,0),\ \text{and}\ (2,-2,0)$, as a function of temperature in samples with long-range (left panel of Fig. 4e) and no (or extremely weak, right panel of Fig. 4f)
CDW order. In previous work as-grown FeGe, we identified a small in-plane lattice distortion around $T_{\rm N}$ and a further lattice expansion at $T_{\rm CDW}$ \cite{wu_symmetry_2024}. We observe a familiar $C6$ rotational symmetry breaking below $T_{\rm N}$ reported previously in the long-range CDW ordered FeGe (left panel of Fig. 4e) \cite{wu_symmetry_2024}. In the low temperature regime $T<150$ K, we observe two spontaneous negative thermal expansions of the kagome lattice at $T_{\rm CDW}$ and $T_{\rm SDW}$ (left panel of Fig. 4f). The lattice expansion at $T_{CDW}$ was reported previously in the as-grown FeGe Ref. \cite{wu_symmetry_2024}, however, we report explicit evidence of coupling between the lattice and SDW order for the first time here.
The right panel of Figure 4f shows the identical measurements for the no (or extremely weak) CDW sample, where there is no observable coupling between the in-plane lattice parameter and SDW order. 

\section{Summary and Conclusions}

In prototypical SDW materials such as the pure metallic Cr, the incommensurate wave vectors of the SDW order are associated with the incommensurate CDW order and both SDW and CDW orders occur below $T_{\rm N}$ \cite{RevModPhys.60.209,RevModPhys.66.25,PhysRevB.51.10336}.For parent compounds of iron pnictide superconductors, the tetragonal-to-orthorhombic lattice distortion and SDW order wave vectors are both commensurate with lattice distortion occurring at temperatures above or at $T_{\rm N}$ \cite{RevModPhys.87.855}. In the case of kagome lattice FeGe, while the CDW transition $T_{\rm CDW}$ occurs below the A-type AFM ordering temperature $T_{\rm N}$, it is above the SDW transition temperature $T_{\rm Canting}$ \cite{doi:10.1143/JPSJ.18.589,beckman_susceptibility_1972,bernhard_magnetic_1988,teng_discovery_2022,PhysRevLett.129.166401,wu_annealing-tunable_2024,WOS:001276353800013,klemm_vacancy-induced_2025}.  Although previous neutron scattering measurements revealed strong evidence for spin-charge-lattice coupling across $T_{\rm CDW}$ \cite{PhysRevLett.133.046502}, a coupling of the lattice to incommensurate magnetic order in FeGe is only observed indirectly through the phonon spectra and electronic bands across the SDW transition \cite{wu_symmetry_2024,oh_tunability_2024}. Our neutron Larmor diffraction data demonstrates a small but sharp distortion in the kagome lattice at the SDW transition $T_{\rm Canting}$  
for the long-range CDW sample (left panel of Fig. 4f), confirming suspicions of a coupling of the lattice and incommensurate magnetic order. No such coupling is found in the no (or extremely weak) CDW sample (right panel of Fig. 4f). As the incommensurate magnetic order of the SDW in FeGe occurs along the $c$-axis and the lattice distortion is within the $ab$-plane, the SDW-lattice coupling in FeGe is much different from that of a typical SDW material like Cr where spin and charge density waves are both incommensurate and occur in the same plane at positions related by a factor of 2 \cite{RevModPhys.60.209,RevModPhys.66.25,PhysRevB.51.10336}.  In a recent X-ray absorption spectroscopy measurements, clear changes are seen in the temperature-dependent Fe $L_{2,3}$-edge X-ray absorption spectroscopy intensity across $T_{\rm CDW}$ and $T_{\rm Canting}$, suggesting the involvement of the orbital degrees of freedom to the CDW and SDW orders \cite{han2024orbitaloriginmagneticmoment}. 

In summary, our neutron results on FeGe samples with and without CDW order show that the conventional understanding of the incommensurate order below $T_{\rm Canting}$ as double-cone AFM order is incorrect for two key reasons. First, we show the persistence of incommensurate spin excitations well above $T_{\rm Canting}$. Second, we identify two distinct spin excitations persisting to at least 2 meV which can not correspond to a simple canting of the A-type AFM order. Rather, 
these results are consistent with
the commensurate gapped excitations arising from the A-type AFM order and the pair of incommensurate gapless excitations arising from SDW order. While the incommensurate order below $T_{\rm Canting}$ is of the SDW type, the SDW-lattice coupling in FeGe is much different from a typical SDW ordered material like Cr \cite{RevModPhys.60.209,RevModPhys.66.25,PhysRevB.51.10336}. Regardless of the microscopic origin of the SDW order, FeGe proves to be a unique system where the energy scales of CDW, SDW, and orbital degrees are very close, and a small disturbance will dramatically affect the transport, magnetic, and electronic properties of the system.  

\begin{acknowledgments}
The neutron scattering and single crystal synthesis work at Rice are supported by US NSF DMR-2401084, DMR-2302420, and the Robert A. Welch Foundation under Grant No. C-1839, respectively (P.D.). M.Y. was supported by the U.S. DOE grant No. DE-SC0021421, the Gordon and Betty Moore Foundation’s EPiQS Initiative through grant No. GBMF9470 and the Robert A. Welch Foundation Grant No. C-2175. This research used resources at the High Flux Isotope Reactor and Spallation Neutron Source, DOE Office of Science User Facilities operated by the Oak Ridge National Laboratory. The beam time was allocated to CORELLI on proposal number IPTS-31981, IPTS-34042.
\end{acknowledgments}

\bibliography{hyspec-bib}

\end{document}